\documentclass[11pt,onecolumn,amssymb,nofootinbib]{revtex4}
\usepackage{amsmath, amsthm, amscd, amssymb}
\usepackage{bm}
\usepackage{bbm}

\begin{document}

\title{\bf Recent Path Crossings with Roman and Anomalies}

\author{Stephen L. Adler}
\email{adler@ias.edu} \affiliation{Institute for Advanced Study,
Einstein Drive, Princeton, NJ 08540, USA.}

\begin{abstract}
I begin with an anecdote, and then discuss my recent work on anomalies in spin-$\frac{3}{2}$  theories.
\end{abstract}

\maketitle
\section{Introduction and an anecdote about Roman}

 The discovery just over 50 years ago of chiral anomalies by me \cite{adler1} and by  Bell and Jackiw \cite{belljackiw} is now featured in a section or chapter  of the standard quantum field theory textbooks, and  my account of work on anomalies during the years from 1968 to 1976 has been written  elsewhere \cite{adler2}, \cite{adler3}.  In this essay I turn to recent interactions with Roman, and with anomalies.  I begin with an anecdote. In 2015 I was invited to attend a workshop held in August at Galiano Island, off Vancouver, on  ``Probing the Mystery: Theory \& Experiment in Quantum Gravity''.  I was scheduled to speak in a session devoted to topics relating to wave function collapse models, as was my friend and occasional collaborator Angelo Bassi.  In discussing with Angelo whether to attend, one of the things that appealed to us, apart from the prospect of good physics and interesting participants, was the fact that Galiano Island is conveniently accessed by seaplane, which intrigued  us both.  So we decided to go.  When we showed up to embark on the seaplane, Roman was also among the group waiting to board.
 When we got onto the plane, Roman turned to me and said  ``I know the plane won't crash; Adler and Jackiw can't die at the same time.''  We did get to the island safely, but the flight was in fact interesting in that there is a high tension line strung across the entrance to the island
 harbor at which we were landing.  To clear this, the seaplane descended to perhaps 10 to 20 feet above the water, and skimmed along this way for quite a distance, passing under the wires, and then settling down into the water.  The procedure was reversed on departure, again the plane
 skimming along just over the water, before climbing to altitude once safely under the wires.  A memorable flight, as was Roman's remark before taking off.

\section{Spin-$\frac{3}{2}$ anomalies and my second recent path crossing with Roman}

My recent encounters with anomalies have all revolved around the question of whether a spin-$\frac{3}{2}$  field can be gauged outside of a supergravity
context, and if it can be, what is its chiral anomaly?  I was led to this by a conjectured grand unified model \cite{adler4} that I have been studying for several years based
on the gauge group $SU(8)$.  The model is suggested in part by rearranging the multiplet content of N=8 supergravity, in a way that preserves boson--fermion
balance (that is equal numbers of boson and fermion degrees of freedom) while not insisting on full supersymmetry.  The fermion content of the
model is one left chiral spin-$\frac{1}{2}$ 56 of $SU(8)$, two left chiral spin-$\frac{1}{2}$ $\overline{28}$ of $SU(8)$, and one left chiral spin-$\frac{3}{2}$ 8 of
$SU(8)$.  If one adopts the standard rule \cite{rule}  that the chiral anomaly of a spin-$\frac{3}{2}$ field in a representation ${\cal R}$ is 3 times the chiral anomaly for a spin-$\frac{1}{2}$ field in the same representation ${\cal R}$, then this representation content cancels $SU(8)$ gauge anomalies \cite{marcus}.

However, in taking this counting seriously, a puzzle arises.  From old work of Johnson and Sudarshan  \cite{js} , and
of Velo and Zwanziger \cite{vz}, it was known that the Dirac bracket for a gauged spin-$\frac{3}{2}$ theory is singular at
small gauge field $\vec B$.  Hence perturbation theory in the gauge coupling breaks down, so how then can
one talk about a one-loop chiral anomaly?  My original thought was that this problem is just a reflection of the fact that when a Rarita-Schwinger field is gauged the free field fermionic gauge invariance is broken.  I found a nice way to restore exact fermionic gauge invariance by adding an auxiliary field \cite{RS1} and hoped  this would improve things. But a careful analysis of the extended gauged model with Henneaux and Pais \cite{ahp} shows that the singularity is just moved to the Dirac bracket for the auxiliary field, and the model remains non-perturbative.

I then noted that in my original unification model \cite{adler4} the group representations allow a direct coupling of the Rarita-Schwinger field to a spin-$\frac{1}{2}$ field, with a coefficient with dimensions of a mass. A simplified Abelian version of this coupling has an interaction term
\begin{equation}\label{coupl}
S_{\rm interaction}=m\int d^4x(\overline{\lambda}\gamma^{\nu}\psi_{\nu}-
\overline{\psi}_{\nu}\gamma^{\nu}\lambda)~~~,
\end{equation}
with $\psi_{\nu}$ the spin-$\frac{3}{2}$ field and $\lambda$ the spin-$\frac{1}{2}$ field. For this model,
one finds that the singular denominator in the Dirac brackets $1/[g \vec \sigma \cdot \vec B(\vec x)]$ is
replaced by $1/[m^2+g\vec \sigma \cdot \vec B(\vec x)]$, with $g$ the gauge coupling. So a perturbation expansion is possible, making it possible to compute the chiral anomaly, and I carried this computation out   in \cite{coupl}.

The anomaly computation for the coupled model is where my second recent path crossing with Roman occurred.
Some of the new textbooks compute chiral anomalies by dimensional regularization, in which the triangle
diagram is kept in its original form with three vertices and three propagators, continued away from
dimension 4.  The problem with using this method in the coupled model is that the propagators and
vertices are complicated matrix structures, and the algebra becomes complicated.  The original  anomaly papers
\cite{adler1} and \cite{belljackiw} noted that the chiral anomaly arises because when applying the zeroth
order Ward identity to the divergence of a current in the triangle, the formal shift of integration
variable needed to get a divergence of zero involves a linearly divergent integral, leading to a nonzero
shift when evaluated carefully.  There is a very nice treatment of the shift method in Roman's lectures
in \cite{rtjg}, which is readily adapted to the  coupled model calculation.   This method has the advantage that after using the zeroth order Ward identity, one
is left with calculating the shift of an expression with only two vertices and two propagators, a considerable simplification.   Carrying
out this computation \cite{coupl} shows that the anomaly contribution from the fermion triangle is
5 times the standard spin-$\frac{1}{2}$ anomaly.  Because the coupling term in Eq. \eqref{coupl} leads to a secondary
constraint with a nonvanishing (and hence Dirac second class) constraint bracket with determinant
${\rm det}[(m^2+g\vec \sigma \cdot \vec B(\vec x))\delta^3(\vec x-\vec y)]$, exponentiation of this
constraint with ghost fields gives a non-propagating ghost with an anomaly contribution of 0.  Hence
the total chiral anomaly in the coupled model remains 5.

Since second class constraints are less familiar than first class ones, I repeated this calculation with
Pais \cite{ap} in the extended coupled model, in which an auxiliary field is added to give an exact
fermionic gauge invariance.  In this case the constraints generate the fermionic gauge transformation, have
vanishing brackets, and so are Dirac first class.  Now the standard Faddeev-Popov method applies; one
adds a gauge fixing action following Nielsen \cite{nielsen}, which gives a ghost contribution to the
chiral anomaly of $-1$.  But when one computes the triangle graph contribution, again following the
shift method in Roman's lectures, one finds an anomaly of 6 times the standard spin-$\frac{1}{2}$ anomaly.  There are other diagrams in the extended model arising from the
auxiliary field construction, and they all give an anomaly contribution of 0. So the total anomaly is 5,
agreeing with the initial calculation I did in \cite{coupl}.

What is interesting about this result is that  it is {\it not} what one  gets by applying the standard lore \cite{rule},
which gives an anomaly of 3 for the spin-$\frac{3}{2}$ field $\psi_\nu$, plus 1 for the spin-$\frac{1}{2}$ field $\lambda$, for a total
anomaly  of 4.  Hence in the only version of Rarita-Schwinger that I can find which admits
a gauging outside of the supergravity context, the standard lore about spin-$\frac{3}{2}$ anomalies gives the
wrong answer!

Because the fermion content of the model of \cite{adler4} was based on the standard anomaly lore, the model
must be modified to cancel $SU(8)$ chiral anomalies.  The natural way to do this is to add a left chiral spin-$\frac{1}{2}$ $\overline{8}$   field to the model.  I am now starting to think about
the implications of doing this.

\section{Happy 80th}

As just described, after a long hiatus of not working on anomalies, my recent studies of spin-$\frac{3}{2}$ brought me back to Roman's excellent lectures on the subject. 
To conclude, very best wishes to Roman on the occasion of his 80th birthday!

\end{document}